\documentstyle[aps,twocolumn,epsfig]{revtex}

\begin{document}

\title{Stability and collective excitations of a two-component 
Bose-condensed gas: \\a moment approach}

\author{Th. Busch${}^{1)}$, J. I. Cirac${}^{1)}$,
        V. M. P\'erez-Garc\'{\i}a${}^{2)}$ and P. Zoller${}^{1)}$}

\address{${}^{1)}$Institut f\"ur Theoretische Physik, Universit\"at Innsbruck, 
A--6020 Innsbruck, AUSTRIA}

\address{${}^{2)}$Departamento de Matem\'aticas, Escuela
 T\'ecnica Superior de Ingenieros Industriales\\
Universidad de Castilla-La Mancha, 13071 Ciudad Real, Spain}

\maketitle

\begin{abstract} The dynamics of a two-component dilute Bose gas of
atoms at zero temperature is described in the mean field approximation
by a two-component Gross-Pitaevskii Equation. We solve this equation
assuming a Gaussian shape for the wavefunction, where the free
parameters of the trial wavefunction are determined using a moment
method. We derive equilibrium states and the phase diagrams for the
stability for positive and negative s-wave scattering
lengths, and obtain the low energy excitation frequencies corresponding to the 
collective motion of the two Bose condensates. 
\end{abstract}


\date{\today}

\narrowtext

\section{Introduction} 

Since the first observations of Bose-Einstein-conden\-sation in an
alkali vapor \cite{Science95,Bradley,Davis} considerable effort has been
made to characterize these systems both experimentally and
theoretically. One of the latest advances is the ex\-pe\-ri\-men\-tal
preparation of two-component Bose-condensates in different internal
states, exhibiting a spatial overlap \cite{Myatt}. In this experiment
${}^{87}$Rb atoms in two different hyperfine states, $|1\rangle =
|1,-1\rangle$ and $|2\rangle = |2,2\rangle$, are loaded into a
magneto-optical trap, are cooled down to effectively zero temperature
and subsequently undergo a phase transition towards a
Bose-Einstein-condensate \cite{Bose}. 

The Gross-Pitaevskii Equation (GPE), a nonlinear Schr\"odinger Equation
(NLSE) for the macroscopic wave function of the Bose condensed gas,
provides an accurate description of the ground state and of the
excitation spectrum of a dilute Bose condensate at zero temperature
\cite{Edwards,Stringari,Baym,Esry,You,Holland,Shly,Castin,Perez,KetterleBoser,KetterleIF,Wallis,Burnett,Ballagh,Stoof}. Recently,
several groups have solved the GPE for single component Bose condensates
and found excellent agreement with experiment \cite{Edwards,Holland}.
The aim of this paper is to study solutions of the {\em two-component}
GPE for a harmonic trapping potential. In particular we will derive the
Bose ground state, investigate its stability properties, and determine
the low-lying excitation frequencies for positive and negative
scattering lengths. Our technique of solving the two-component GPE is
based on a Gaussian ansatz for the condensate wave function where the
open parameters are determined by moment methods \cite{nonlineq}
(which is equivalent to
a variational technique \cite{Perez}). This allows one to obtain
essentially analytical solutions of the GPE, and thus complements
numerical studies of this equation \cite{EsryGreen}. 

The paper is organized as follows. In Sect. 2 of this paper we first
define our model, and derive the equations of motion for the parameters
of our Gaussian wave function. In particular, we study in detail the
simple case of of isotropic traps and isotropic condensates. In Sec. 3
we investigate first the case where the centers of the two trapping
potentials for the two condensates coincide. For this configuration we
calculate the equilibrium points and the low energy eigenmodes. In
addition, we analyze the stability of the system. Finally, in Sec. 4 we
consider the case of displaced trap centers, and compare with the
results of Sec. 3.

\section{Basic equations}
\subsection{The model}

Zero-temperature mean-field theory provides an accurate theoretical 
description of dilute Bose condensed systems. In particular, the dynamics 
of a two-component Bose gas can be modeled by the coupled 
Gross-Pi\-taeves\-kii equations 
\begin{mathletters}
\label{NLSE}
\begin{eqnarray}
 i\partial_t \Psi_1 &=& \left[ 
     - \frac{1}{2M}\nabla^2
     + V_1(\vec r)
     +\sum_{n=1}^2 U_{1n} |\Psi_n|^2\right] \Psi_1,\\
 i\partial_t \Psi_2 &=& \left[ 
     - \frac{1}{2M}\nabla^2
     + V_2(\vec r)
     +\sum_{n=1}^2 U_{2n} |\Psi_n|^2\right] \Psi_2
\end{eqnarray}
\end{mathletters}
which are a set of nonlinear Schr\"odinger equations for the macroscopic wave 
functions 
$\Psi_1$ and $\Psi_2$ for the two components of the condensate. In 
Eqs.~(\ref{NLSE}) $V_n$,$(n=1,2)$ denotes the trap potentials which we 
assume to be harmonic in agreement with common experimental situations
\begin{equation}
 V_n(\vec r)=\frac{1}{2}M \omega^2 \left( \lambda_x^2 (x-x_{n})^2 
             +\lambda_y^2 y^2 + \lambda_z^2 z^2\right). 
\end{equation} 
The parameters $\lambda_\eta\; (\eta=x,y,z)$ account for the 
an\-iso\-tro\-py of the trap, and the trap centers for the first and se\-cond 
component are displaced in the $x$-direction by $x_1$ and $x_2$, 
respectively. The  coupling constants $U_{nm}$ are related to the scattering 
lengths $a_{nm}$ by $U_{nm}=4\pi\hbar^2 a_{nm}/M$. 
In this paper we allow only for the following elastic scattering processes:
$|1\rangle |1\rangle \rightarrow |1\rangle |1\rangle $,
$|2\rangle |2\rangle \rightarrow |2\rangle |2\rangle $,
$|1\rangle |2\rangle \rightarrow |1\rangle |2\rangle $.
Up to now, Eqs.~(\ref{NLSE}) have been solved in the Thomas-Fermi 
approximation, exploring the spatial densities of the mixtures \cite{Ho} 
and the long wavelength excitations \cite{Graham}. Numerical studies of 
these systems have been carried out in \cite{EsryGreen}.

For non-interacting particles the Eqs.~(\ref{NLSE}) are solved for the 
ground state by Gaussian functions, which are no more solutions in the 
general case due to the nonlinear self--interaction and coupling terms.
The basic assumption behind the following derivations is that, 
in the case of interactions, the wave function is still well approximated by 
a Gaussian,
\begin{equation}
\label{WF}
\Psi_n(\vec r,t)= A_n(t) e^{-\frac{1}{2}\left[\beta_{nx}(t) (x-x_{n0})^2 +
                                              \beta_{ny}(t) y^2 +
                                              \beta_{nz}(t) z^2
                                        \right]}
\end{equation}
where the $\beta_{n\eta}=\beta_{n\eta}^R + i \beta_{n\eta}^I$ are complex
numbers. The adjustable parameters in it can be interpreted as the
amplitudes $A_n$, 
the width $\text{w}_{n\eta}=1/\sqrt{\beta_{n\eta}^R}$ and the curvature 
$(M_{n\eta}\sqrt{\beta_{n\eta}^R})^{-1/2}=(\beta_{n\eta}^I)^{-1/2}$ of the
wavefunction. To derive reliable results for the corresponding equations 
of motion it is necessary to take also the imaginary part of the wavefunction
into account 
as shown in \cite{Desaix} for a similar case. One has to keep in mind 
when applying this ansatz to problems where losses are included, the loss 
terms usually induce 
aberrations, thus making the profile non--Gaussian. When nonlinear
loss or gain terms are substantially important these aberrations 
make the information that can be obtained form an ansatz like Eqs. (\ref{WF})
only qualitative. However, in the present case when losses are small 
corrections and do not determine the dynamics the solutions of 
Eq. (\ref{NLSE}) can be well approximated by Gaussian functions.

In the following we find it convenient to work with scaled variables. We 
will measure energies in units of  $\hbar \omega$, and will scale lengths 
with respect to $a_0 \sqrt{2}$ where $a_0=[\hbar/(M\nu)]^{1/2}$  
is the size of the ground state of the bare harmonic oscillator.

\subsection{Moment equations}

In Ref. \cite{Perez} solution of the  single component Gross-Pitaevskii 
equation was formulated as a variational problem, where a Lagrangian density 
$\cal L$ is derived so that the stationary point of the corresponding action 
gives the nonlinear Schr\"odinger equation.  An approximate solution of the 
GPE was then derived by finding the extremum of the action within a set
of Gaussian trial functions which give ``Newton type'' equations for the 
variational parameters. In contrast to such a  variational ana\-ly\-sis , 
we will 
solve  the two-component GPE by means of a  moment-method. In the case of 
real $U_{nm}$, i.e. no losses \cite{losses}, this method is equivalent to the 
variational approach. In principle, this approach is more general, as it can 
also be applied in a situation, where loss terms are included in 
Eqs.~(\ref{NLSE}).

Using the Gaussian ansatz for a system where one condensate is centered at
$x_1=0$ and the second one at $x_2=\alpha$ it is possible to calculate the 
number of particles in each condensate  $n=1,2$ as a function of the parameters
\begin{equation} \label{Nn}
 N_n=\int_{-\infty}^{\infty}d^3\vec r\;|\Psi_n(\vec r,t)|^2
    =|A_n(t)|^2\pi^{3/2}\text{w}_{nx}\text{w}_{ny}\text{w}_{nz} \; .
\end{equation}
In a similar way, the second moments of the spatial coordinates $\eta=x,y,z$ 
and the second moments of the momenta  are given by
\begin{mathletters}
\label{eta}
\begin{eqnarray}
\langle\eta^2\rangle_1
                 &=&\int_{\infty}^{\infty}d^3\vec r\;\eta^2|\Psi_1(\vec r,t)|^2
                  =\frac{1}{2}N_1\text{w}_{1\eta}^2,\\
\langle\eta^2\rangle_2
                 &=&\int_{\infty}^{\infty}d^3\vec r\;\eta^2|\Psi_2(\vec r,t)|^2
                  =\frac{1}{2}N_2\text{w}_{2\eta}^2
                    (1+\frac{2\alpha}{\text{w}_{2\eta}^2})
\end{eqnarray}
\end{mathletters}
and
\begin{eqnarray} \label{deriv}
\langle \partial_\eta^2\rangle_n&=&\int_{\infty}^{\infty}d^3\vec r\;
                              \Psi_n(\vec r,t)^\ast\partial_\eta^2\Psi_n(\vec
r,t)\nonumber\\
                          &=&-\frac{N_n}{2}(\text{w}_{n\eta}^{-2}+M_{n\eta}^2)
\end{eqnarray}
respectively. These fourteen variables provide a complete description of the 
system within the previously mentioned approximation. To obtain a closed set 
of equations of 
motion for $N_n(t)$, $\text{w}_{n\eta}(t)$ and $M_{n\eta}(t)$, we take the 
time derivative of Eqs.~(\ref{Nn},\ref{eta},\ref{deriv}), eliminate the time 
derivatives of the wave function with the help of the Schr\"odinger equations 
(\ref{NLSE}), and work out the resulting integrals over $\vec r$ with the 
help of the Gaussian ansatz (\ref{WF}). We obtain 

\begin{mathletters}
\begin{eqnarray}
\dot N_1 &=& \frac{2 q^I_{11} N_1^2}{\text{w}_{1x}\text{w}_{1y}\text{w}_{1z}}
             +2 \sqrt{8} q_{12}^I N_1 N_2                    
             e^{-\frac{\alpha^2}{\text{w}_{1x}^2+\text{w}_{2x}^2}}\times\nonumber\\ 
          && \left[(\text{w}_{1x}^2 + \text{w}_{2x}^2)
                   (\text{w}_{1y}^2 + \text{w}_{2y}^2)
                   (\text{w}_{1z}^2 + \text{w}_{2z}^2)\right]^{-\frac{1}{2}},\\
\dot{\text{w}}_{1x} &=&-M_{1x}-\frac{N_1q_{11}^I}{2\text{w}_{1y}\text{w}_{1z}}
                       +\sqrt{8}q_{12}^I N_2 \times\nonumber\\
          && \left[(\text{w}_{1x}^2 + \text{w}_{2x}^2)
                   (\text{w}_{1y}^2 + \text{w}_{2y}^2)
    (\text{w}_{1z}^2 + \text{w}_{2z}^2)\right]^{-\frac{1}{2}}\times\nonumber\\
          && \frac{\text{w}_{1x}^3}{(\text{w}_{1x}^2+\text{w}_{2x}^2)^2}
             [\text{w}_{1x}^2-\text{w}_{2x}^2-2\alpha^2]
          e^{-\frac{\alpha^2}{\text{w}_{1x}^2+\text{w}_{2x}^2}},\\
\dot{\text{w}}_{1y} &=&-M_{1y}-\frac{N_1q_{11}^I}{2\text{w}_{1x}\text{w}_{1z}}
                       +\sqrt{8}q_{12}^I N_2 \times\nonumber\\
          && \left[(\text{w}_{1x}^2 + \text{w}_{2x}^2)
                   (\text{w}_{1y}^2 + \text{w}_{2y}^2)
    (\text{w}_{1z}^2 + \text{w}_{2z}^2)\right]^{-\frac{1}{2}}\times\nonumber\\
          && \frac{\text{w}_{1y}^3}{\text{w}_{1y}^2+\text{w}_{2y}^2}
          e^{-\frac{\alpha^2}{\text{w}_{1x}^2+\text{w}_{2x}^2}},\\
\dot{M}_{1x} &=& \text{w}_{1x}\lambda_x^2-\frac{1}{\text{w}_{1x}^3}
                 -\frac{N_1}{\text{w}_{1x}\text{w}_{1y}\text{w}_{1z}}
         [\frac{3}{2}q_{11}^IM_{1x}+\frac{q_{11}^R}{\text{w}_{1x}}]\nonumber\\
         &&+\sqrt{8}N_2
          \left[(\text{w}_{1x}^2 + \text{w}_{2x}^2)
                (\text{w}_{1y}^2 + \text{w}_{2y}^2)
   (\text{w}_{1z}^2 + \text{w}_{2z}^2)\right]^{-\frac{1}{2}}\times\nonumber\\
                   &&\frac{1}{(\text{w}_{1x}^2+\text{w}_{2x}^2)^2}
                   \Big[q_{12}^I(\frac{2}{M_{1x}}
                   (\text{w}_{1x}^2+\text{w}_{2x}^2-2\alpha^2)-\Big.\Big.\nonumber\\
&&\Big.\Big.M_{1x}(\text{w}_{1x}^4+2\text{w}_{2x}^4+3\text{w}_{1x}^2\text{w}_{2x}^2+2\alpha^2\text{w}_{1x}^2)\Big)\Big.\nonumber\\
&&-2q_{12}^R\text{w}_{1x}[\text{w}_{1x}^2+\text{w}_{2x}^2-2\alpha^2]\Big]
e^{-\frac{\alpha^2}{\text{w}_{1x}^2+\text{w}_{2x}^2}},\\
\dot{M}_{1y} &=& \text{w}_{1y}\lambda_y^2-\frac{1}{\text{w}_{1y}^3}
                 -\frac{N_1}{\text{w}_{1x}\text{w}_{1y}\text{w}_{1z}}
         [\frac{3}{2}q_{11}^IM_{1y}+\frac{q_{11}^R}{\text{w}_{1y}}]\nonumber\\
         &&+\sqrt{8}N_2
          \left[(\text{w}_{1x}^2 + \text{w}_{2x}^2)
                (\text{w}_{1y}^2 + \text{w}_{2y}^2)
   (\text{w}_{1z}^2 + \text{w}_{2z}^2)\right]^{-\frac{1}{2}}\times\nonumber\\
                   &&\frac{1}{(\text{w}_{1x}^2+\text{w}_{2x}^2)^2}
                   \Big[q_{12}^I(\frac{2}{M_{1x}}
                   (\text{w}_{1x}^2+\text{w}_{2x}^2)-\Big.\Big.\nonumber\\
&&\Big.\Big.M_{1x}(\text{w}_{1x}^4+2\text{w}_{2x}^4+3\text{w}_{1x}^2\text{w}_{2x}^2)\Big)\Big.\nonumber\\
&&-2q_{12}^R\text{w}_{1x}[\text{w}_{1x}^2+\text{w}_{2x}^2]\Big]
e^{-\frac{\alpha^2}{\text{w}_{1x}^2+\text{w}_{2x}^2}},
\end{eqnarray}
\end{mathletters}
where we replaced $q_{nm}=U_{nm}{\sqrt{8\pi^3}}$ \cite{losses}. 
The corresponding equations 
for the second condensate are obtained by exchanging every index $1$ and $2$ 
and the corresponding equations for the $z$-direction are obtained by 
exchanging the indices $y$ and $z$ in the equations for the $y$-direction. 

\section{Isotropic traps and identical condensates: no displacement of trap centers}

\subsection{Equations}

The simplest case which is amenable to analytical treatment corresponds to an
isotropic trapping potential $\lambda_x=\lambda_y=\lambda_z=1$ with 
(non-displaced) trap centers $x_n =0$ ($n=1,2$)
\cite{noteexp}. 
We will study the case of a fixed number of particles in
each condensate, $U_{11}^I=U_{22}^I=U_{12}^I=0$, and, in particular, 
consider them equal $N_1=N_2=N$. With these 
assumptions, and the initial condition of two isotropic condensates
$\text{w}_{nx}=\text{w}_{ny}=\text{w}_{nz} \equiv \text{w}_n$ ($n=1,2$),
the equations of motion for the widths of the condensates have the form
\begin{equation} \label{dot}
 \ddot{\text{w}}_1=-\text{w}_1+\frac{1}{\text{w}_1^3}
                   +\frac{Nq_{11}}{\text{w}_1^4}
+2\sqrt{8}\frac{Nq_{12}\text{w}_1}{(\text{w}_1^2+\text{w}_2^2)^{\frac{5}{2}}}.
\end{equation}
whereby exchanging the indices we get the corresponding one for $\text{w}_2$.

Similar to Ref. \cite{Perez} we can find a ``potential'' for these equations.
Hence, the evolution of the width of the condensates can be viewed as the 
coordinates
of a fictitious classical particle moving in a three dimensional potential 
\begin{eqnarray}
\label{potential}
V_{\text{eff}}(\text{w}_1,\text{w}_2)&=&\frac{1}{2}(\text{w}_1^2+\text{w}_2^2)
                                 +\frac{1}{2}\left(\frac{1}{\text{w}_1^2}
+\frac{1}{\text{w}_2^2}\right) 
                                 \nonumber\\
 &&+\frac{1}{3}N
    \left(\frac{q_{11}}{\text{w}_1}+\frac{q_{22}}{\text{w}_2}\right)
+\frac{2}{3}\sqrt{8}\frac{Nq_{12}}{(\text{w}_1^2+\text{w}_2^2)^{\frac{3}{2
}}}
                                 \nonumber\\
\end{eqnarray}
\\
where the first term on the RHS stems from the harmonic trapping potential, 
the second term results from the kinetic energy term, and the third and 
fourth term are due to the interaction between alike particles and between 
particles in different states.
Below we further reduce the number of open parameters by assuming the  
scattering lengths $a_{11}$ and $a_{22}$ to be equal.

\subsection{Equilibrium Points}

If we now examine the state in equilibrium of the system, we find that the
requirement of equal scattering lengths implies that the condensates must
have equal widths $\text{w}_1=\text{w}_2 \equiv\text{w}$. Equilibrium points 
are obtained from the  extrema of the potential (\ref{potential}), which gives 
\begin{equation}
\label{eom_eq}
 \text{w}=\frac{1}{\text{w}^3}+\frac{N}{\text{w}^4}(q_{11}+q_{12})  .
\end{equation}

In the hydrodynamic approximation this equation of motion reduces further and 
can be analytically solved as 
$\text{w}_{eq}^{ha}=N^{1/5}(q_{11}+q_{12})^{1/5}$.
To test for  sta\-bi\-li\-ty of these equilibrium points requires us to 
check for minima with the help of the Hessian-\-matrix 
$H=[\partial^2 V_{\text{eff}}/(\partial\text{w}_{1n}\partial\text{w}_{2n}
)]_{\text{w}_{eq}}$. This analysis will be carried out in the following 
subsection.

\subsection{Low energy eigenmodes and Stability}

To obtain the frequencies of small amplitude oscillations around the 
equilibrium positions we linearize the full equations (\ref{dot}) around the 
equilibrium points, 
$\text{w}_{n\eta}=\text{w}_{n\eta}^{eq}+\delta\text{w}_{n\eta}$. 
We find the following four frequencies:
\begin{mathletters}
\label{modes}
\begin{eqnarray} 
&\omega_{a}&=\sqrt{1+\frac{3}{\text{w}^4}+\frac{N}{\text{w}^5}(q_{11}-q_{1
2})}\\
&\omega_{b}&=\sqrt{1+\frac{3}{\text{w}^4}+\frac{N}{\text{w}^5}(q_{11}+q_{1
2})}\\
&\omega_c&=\sqrt{1+\frac{3}{\text{w}^4}+\frac{N}{\text{w}^5}(4q_{11}+4q_{12})}\\
&\omega_d&=\sqrt{1+\frac{3}{\text{w}^4}+\frac{N}{\text{w}^5}(4q_{11}-q_{12})
}
\end{eqnarray}
\end{mathletters}
which can be interpreted as collective oscillations depicted in  Fig.~1. The 
modes $(a)$ and $(b)$ are two fold degenerate, corresponding to oscillation 
e.g. in the $x-y$-plane and the $x-z$-plane \cite{anothernote}. Modes  $(c)$ 
and $(d)$ corres\-pond to isotropic oscillations.
\begin{figure}
\begin{center}
  \epsfig{file=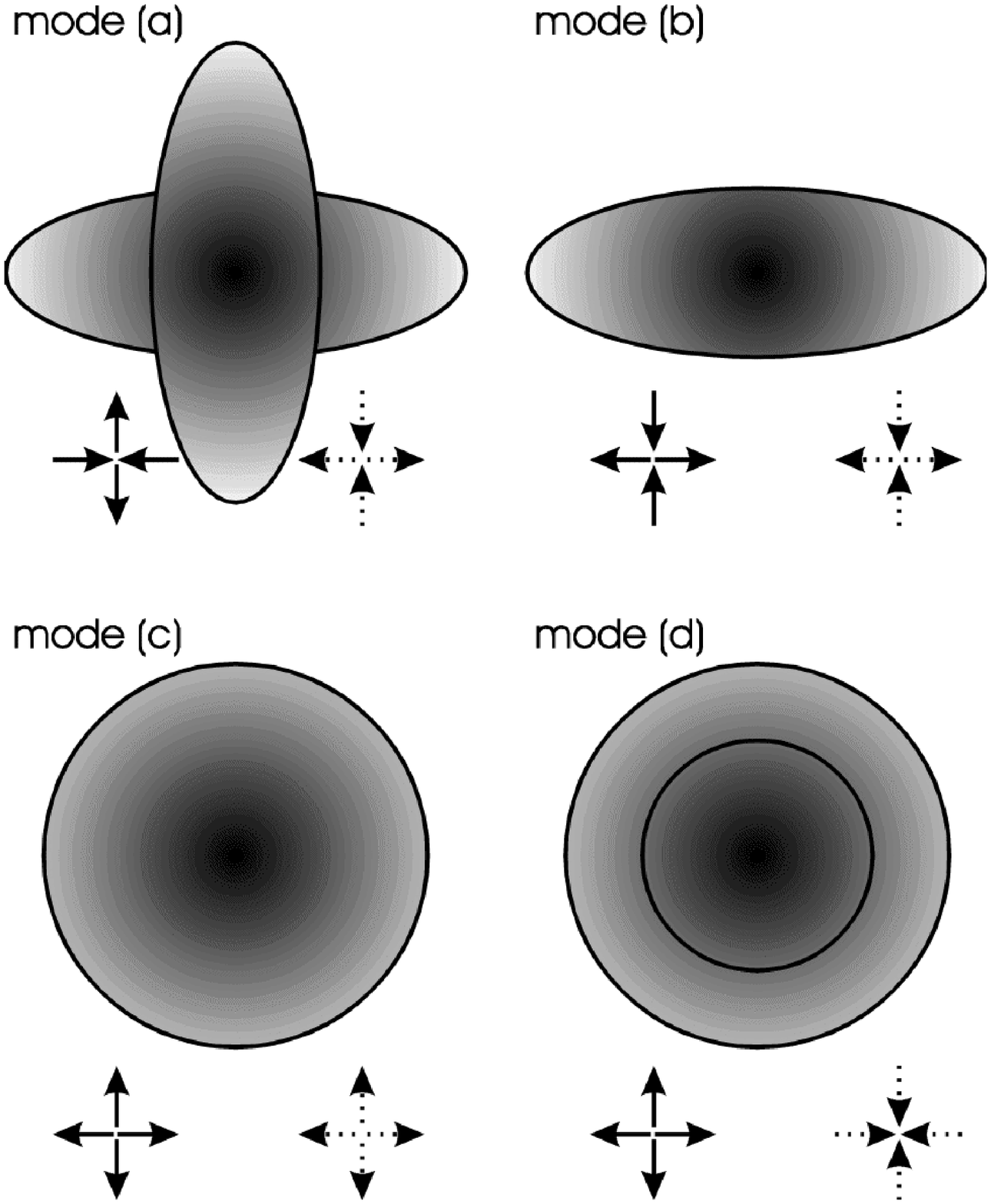,width=\linewidth}
  \begin{caption}
{\sf 
Diagrammatical representation in the $x$-$y$-plane of the four collective 
excitations with the lowest energy. Oscillations of particles in state 
$|1\rangle$ are visualized by solid line arrows, oscillations of particles in 
state $|2\rangle$ by dashed line arrows.
}
  \end{caption}
\end{center}
\end{figure}
The order in which these motions appear with increasing energy depends 
strongly on the values of the scattering lengths involved.  In Fig.~2 we
display the spectrum for a positive value of $a_{11}$. For large, negative 
values of $a_{12}$, the two modes with the lowest energies are those with
the maximum spatial overlap, namely mode $b$ and mode $c$.  If we now 
consider the region where $a_{12}$ is positive and large, we naturally find 
that the modes $a$ and $d$, for which the spatial overlap is minimized, are 
the lowest energy modes. The point at which this crossing happens changes with
changing values of $a_{11}$. Again we find, as in \cite{Perez}, that the 
widths of the condensates remain finite up to the point of  collapse.
\begin{figure}
\begin{center}
  \epsfig{file=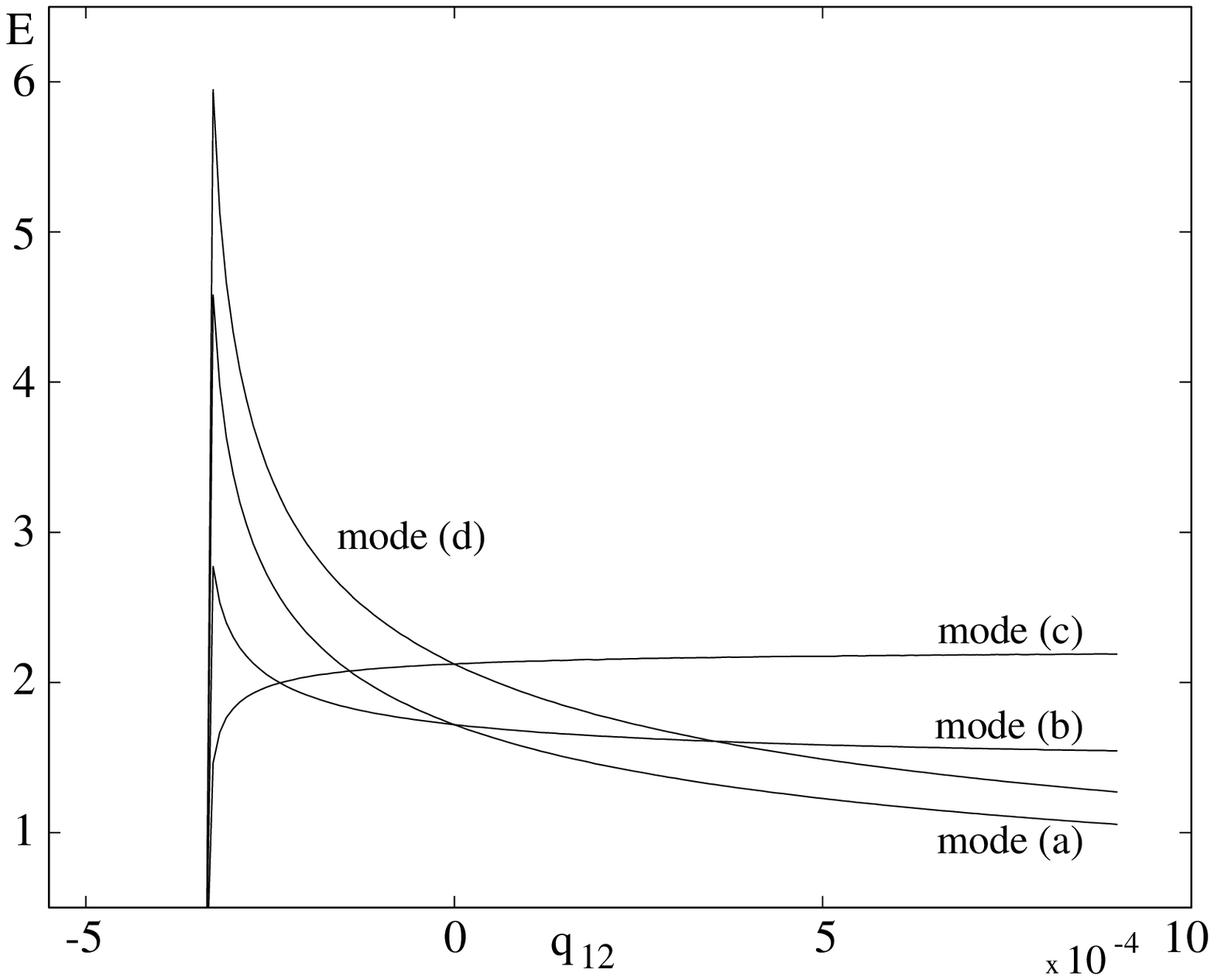,width=\linewidth,height=6cm}
  \begin{caption}
{\sf Spectrum of the low lying excitation frequencies for a fixed, positive 
value $q_{11}=2.3\times 10^{-4}$. With $q_{12}$ increasing from negative 
values to positive ones the two low lying modes and the two high lying ones 
get exchanged.
}
  \end{caption}
\end{center}
\end{figure}
The eigenfrequencies (\ref{modes}) provide the stability region of the 
equilibrium points (\ref{eom_eq}),
\begin{mathletters}
\begin{eqnarray}
 N(q_{11}+q_{12})&=&-\frac{4}{5}\left(\frac{1}{5}\right)^{\frac{1}{4}}\\
  -\left(\frac{5}{4}\right)^{5}N^4q_{11}^5+\frac{1}{4}q_{11}&=&q_{12}.
\end{eqnarray}
\end{mathletters}
Since for $a_{12}=0$ we recover the single component results 
\cite{Perez} where only
particles in one state are present, we see that for positive 
values of $a_{12}$ the region of stability increases and for negative ones it 
decreases. This is due to contributions to the total system energy from either
the repulsive or attractive interaction. However, this effect is non-linear 
due to the nature of this interaction. It is illustrated in Fig.~3.
\begin{figure}
\begin{center}
  \epsfig{file=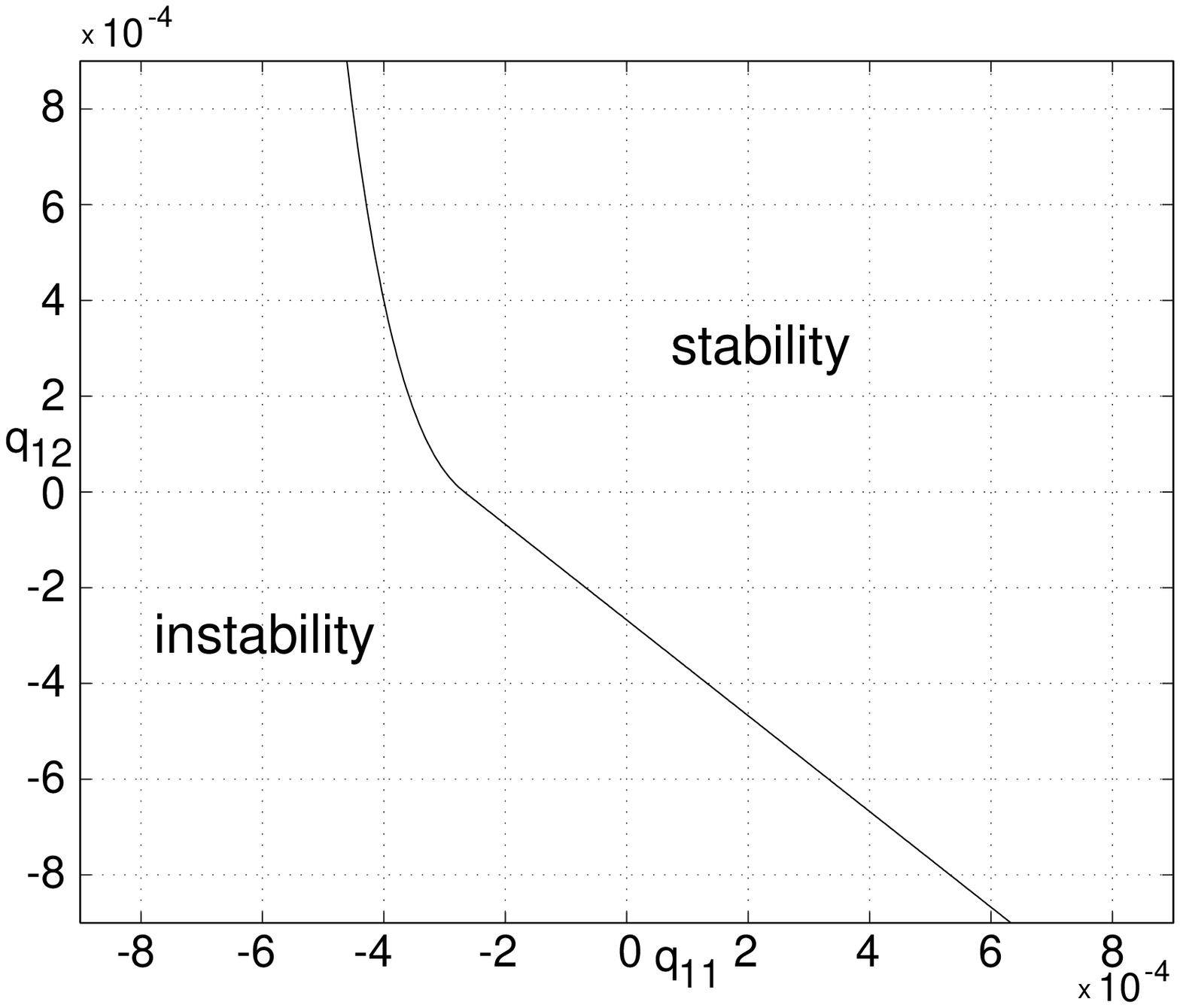,width=\linewidth,height=6cm}
  \begin{caption}
{\sf 
Phase diagram for the stability of the two-condensate system for $N=2000$ 
particles. For positive $a_{12}$ the region of stability is increased due to 
the positive energy contribution from the repulsive interaction. However,
it does not expand linearly with increasing $a_{12}$. 
}
  \end{caption}
\end{center}
\end{figure}

\section{isotropic traps and identical condensates: displaced trap centers}

\subsection{Equations}
Let us now consider the case of two spatially displaced traps and work
out the differences to the above situation. We assume a displacement in the
$x$-direction and make 
an ansatz analogous to Eqs.(\ref{WF})  and (\ref{NLSE}) where one condensate 
is centered at 
$x_{1}=0$ and the second at $x_{2}=\alpha$. Also in this case we
assume that the number of particles is conserved, what implies that the 
imaginary parts of the coupling constants are again zero. We apply the 
moment method as 
in the previous case and arrive, since the spatial symmetry is now broken in 
the $x$-direction, at the following set of equations of motion for the widths 
of the two condensates in the $x$-direction, $\text{w}_{nx}$, and in the $y$- 
and $z$-directions, $\text{w}_n$ 
\begin{mathletters}
\label{dteom}
\begin{eqnarray} 
\ddot{\text{w}}_{1x}&=&-\text{w}_{1x}\lambda_{1x}^2+\frac{1}{\text{w}_{1x}^3}
+\frac{N_1q_{11}}{\text{w}_{1x}^2\text{w}_1^2}\nonumber\\
                     &&\hspace*{-0.4cm}+2\sqrt{8}N_2
                         \frac{q_{12}\text{w}_{1x}(\text{w}_{1x}^2
                                                  +\text{w}_{2x}^2      
                                                  -2\alpha^2)}
                              {(\text{w}_{1x}^2+\text{w}_{2x}^2)^{\frac{5}{2}}
                               (\text{w}_1^2+\text{w}_2^2)}
                        e^{-\frac{\alpha^2}{\text{w}_{1x}^2+\text{w}_{2x}^2}}\\
\ddot{\text{w}}_1&=&-\text{w}_1\lambda_1^2+\frac{1}{\text{w}_1^3}
                       +\frac{N_1q_{11}}{\text{w}_{1x}\text{w}_1^3}\nonumber\\
                     &&\hspace*{-0.4cm}+2\sqrt{8}N_2
                         \frac{q_{12}\text{w}_1}
                              {(\text{w}_{1x}^2+\text{w}_{2x}^2)^{\frac{1}{2}}
                               (\text{w}_1^2+\text{w}_2^2)^2}
                        e^{-\frac{\alpha^2}{\text{w}_{1x}^2+\text{w}_{2x}^2}}.
\end{eqnarray}
\end{mathletters}
Again one gets the equations for $\text{w}_{2x}$ and $\text{w}_2$ by simply
exchanging the indices. We can also like, in the non-displaced case, find a 
potential of the form of Eq.(\ref{potential}), which is suitable for the 
derivation of these equations.  However, in this case the term originating 
from the interaction of the particles in the different states is multiplied by
an additional factor, $\exp[-\alpha^2/(\text{w}_{1x}^2+\text{w}_{2x}^2)]$, 
which ensures that the influence of these interactions gets weaker, the more 
the condensates are displaced.  This then implies an even smaller region of 
spatial overlap. Note that the widths of the condensates are included in this 
factor and it therefore has a non-negligible effect on the physics of the 
system in addition to weakening the magnitude of effects found in the 
non-displaced case.\\
Let us start calculating the equilibrium states for this case from 
Eqs.(\ref{dteom}). In Fig.~4 we plotted the widths $\text{w}_x$ and 
$\text{w}$ of the condensates for $a_{11}$ and $a_{12}$ positive and equal. 
Once the condensates are separated by a distance 
$\alpha=\sqrt{(\text{w}_{1x}^2+\text{w}_{2x})/2}$ the repulsive interaction 
between particles from different states becomes an effectively attractive one 
(and v.v.), as can be seen from the fact that the width on the $x$-direction 
gets smaller then the width in the $y$- and $z$-direction. The explanation for
this effect is, that the direction of the repulsive force with respect to the 
surfaces of the condensates changes in these regions where they do no longer 
overlap. At large distances when the interaction between the particles in 
different states becomes negligible again, the case of two independent, 
isotropic condensates is recovered again.
\begin{figure}
\begin{center}
  \epsfig{file=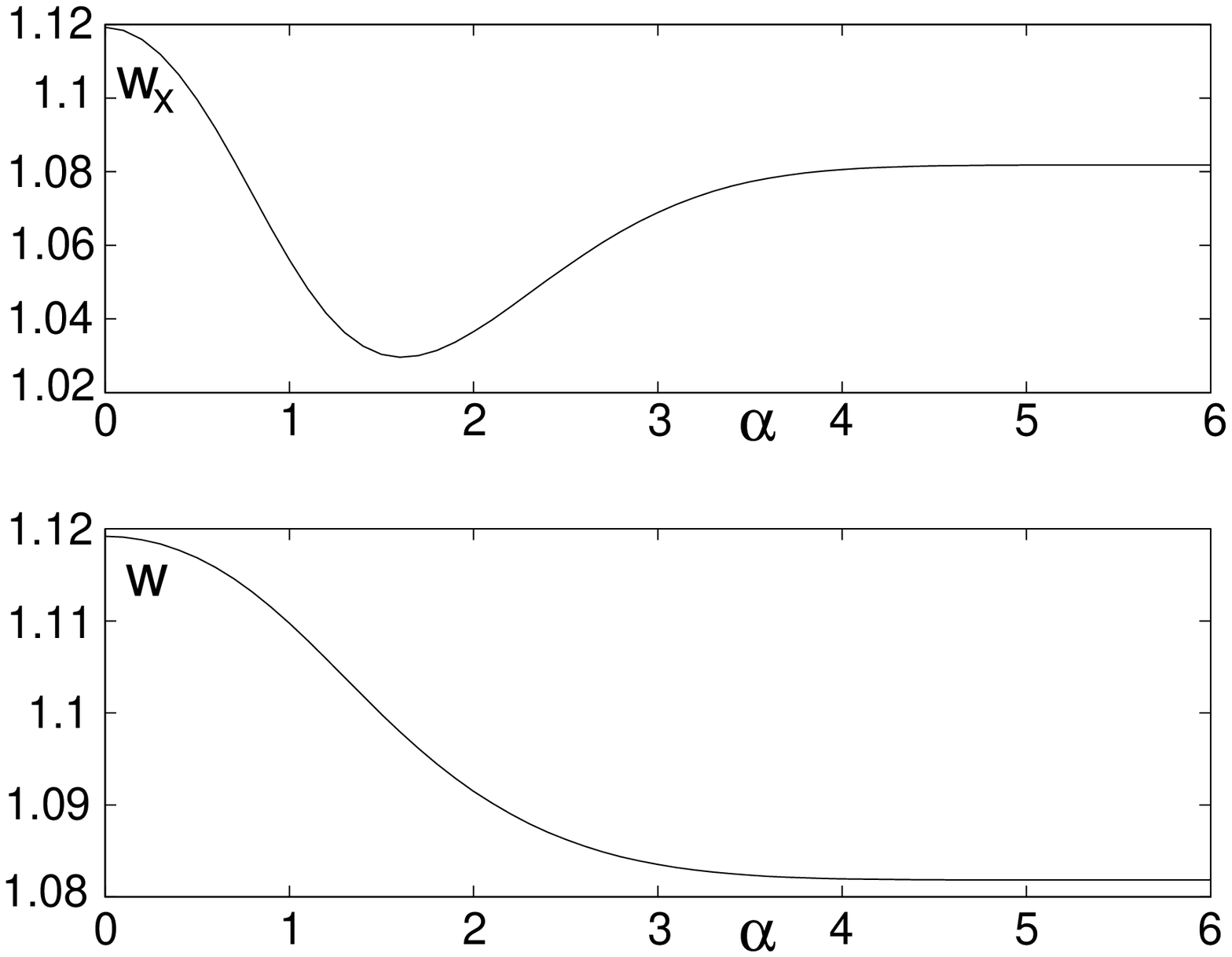,width=\linewidth,height=6cm}
  \begin{caption}
{\sf 
Width of the condensates in the $x$-direction and in the $y$-,$z$-directions 
for $q_{11}=0.2\times 10^4$ and $q_{12}=0.2\times 10^4$.  Due to the repulsive 
interaction between particles from different states, the condensate gets 
squeezed in the $x$-direction at a certain distance and then expands once
the region of overlap diminishes further. 
}
  \end{caption}
\end{center}
\end{figure}
\subsection{Low energy eigenmodes and Stability}
Looking at the excitation spectrum of the two component system by analyzing it
again via a linear expansion around the equilibrium points, we find, as can be
seen from Fig.~5, that the degeneracies existing in the non-displaced case 
get lifted with increasing distance. Bringing the condensates that far from 
each other that they can be considered as two independent condensates 
\cite{Perez} the spectrum reduces again to two three-fold degenerate values, 
since the condensates become again isotropic.

\begin{figure}
\begin{center}
  \epsfig{file=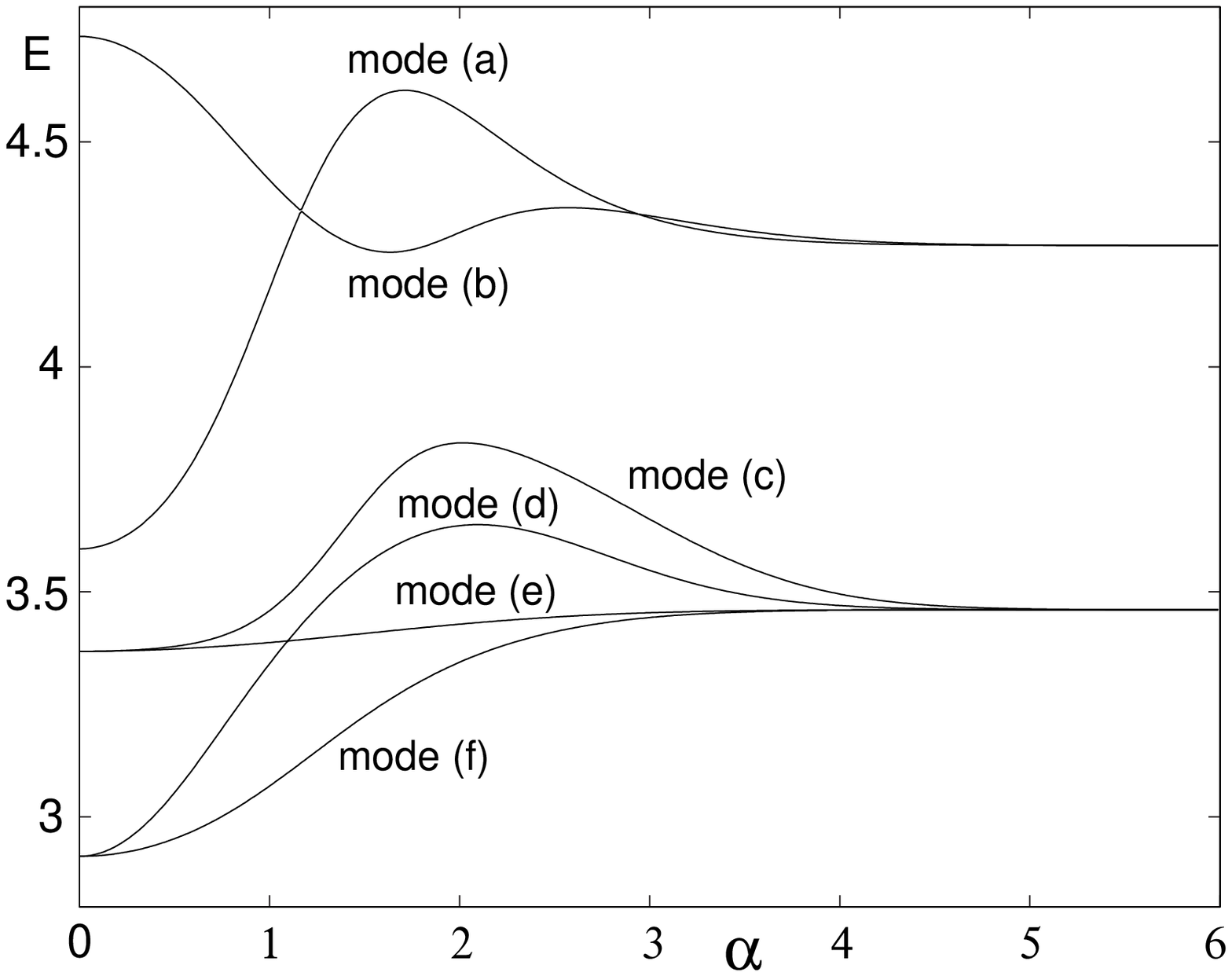,width=\linewidth,height=6cm}
  \begin{caption}
{\sf 
Spectrum of the low lying excitation energies for 
$q_{11}=q_{12}=6\times 10^{-4}$. The degeneracies are lifted with increasing
$\alpha$ and for $\alpha>2$ it reduces to the case of two ,isotropic, single 
condenstates again.
}
  \end{caption}
\end{center}
\end{figure}

In this Gaussian-ansatz we can again easily interpret these modes to be of the 
following types:
\begin{equation}
\begin{array}{rrrrrrrl}
(&-\gamma_a&-1&-1& \gamma_a& 1& 1)&\hspace{0.5cm}\text{mode(a)}\nonumber  \\ 
(& \gamma_b& 1& 1& \gamma_b& 1& 1)&\hspace{0.5cm}\text{mode(b)}\nonumber   \\
(& \gamma_c&-1&-1& \gamma_c&-1&-1)&\hspace{0.5cm}\text{mode(c)}\nonumber  \\
(&-\gamma_d& 1& 1& \gamma_d&-1&-1)&\hspace{0.5cm}\text{mode(d)}\nonumber  \\
(&0&-1& 1& 0&-1& 1)&\hspace{0.5cm} \text{mode(e)}\nonumber  \\
(&0& 1&-1& 0&-1& 1)&\hspace{0.5cm} \text{mode(f)}\nonumber  \\
\end{array}
\end{equation}
The members of these vectors represent the oscillation-amplitude in the
directions 
$(\delta_{1x}\;\delta_{1y}\;\delta_{1z}\;\delta_{2x}\;\delta_{2y}\;\delta_{2z})$, 
where $\gamma_l\;(l=a,b,c,d)$ depends on the parameters of the system.
The lifting of the degeneracies is mainly due to the fact, that the spatial
displaced system is not invariant under an exchange of the variables  
$\text{w}_{1y}$ and $\text{w}_{1z}$ and not of
$\text{w}_{2y}$ and $\text{w}_{2z}$, as can be seen from Eqs. (\ref{dteom}).

Let us  now look at the stability of the stationary states, which we again
retrieve from the analysis of the Hessian matrix. As can be seen from Fig.~6 
the further the condensates separate from each other, the weaker the effect 
of additional stabilization (destabilization) becomes, since it results from  
the repulsive (attractive) interaction between the condensates.  For 
$\alpha\approx> 1.5$ one recovers the stability condition for the case 
of two separate condensates. 
\begin{figure}
\begin{center}
  \epsfig{file=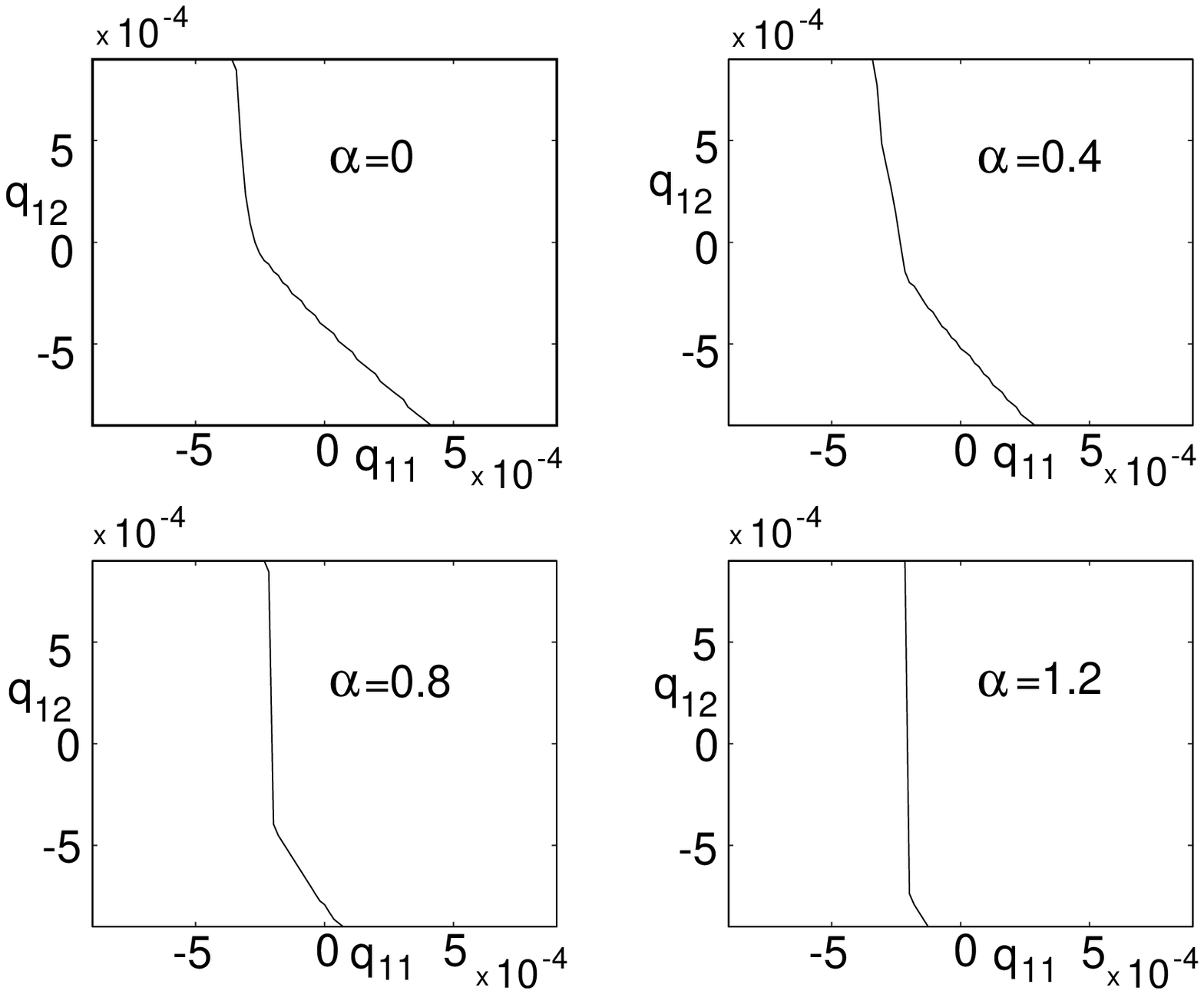,width=\linewidth,height=6cm}
  \begin{caption}
{\sf 
Stability region for displaced traps with a displacement parameter $\alpha$. 
For far displaced traps the result for independent condensates is recovered.
}
  \end{caption}
\end{center}
\end{figure}

\section{conclusions}
We used a moment method to describe the behavior of two overlapping
Bose-Einstein-Condensates in different internal states, assuming that the 
shapes of the individual wave functions do not deviate too much from a 
Gaussian-shape. With this method it is straightforward to derive equations 
of motion for the open parameters defining the exact form of the Gaussian 
function, especially the width of the condensates. They can for
the general cases easily treated numerically and for special cases even 
analytical solutions are possible.

For the case of isotropic traps and isotropic condensates we did calculate the
equilibrium points for the system and analyzed their stability. We found, 
that in comparison to the case of one pure condensate the region of stability 
can be extended (shrinked) if the interaction energy between particles in 
different states is positive (negative). Due to the nonlinear character of
the interaction also this effect is of nonlinear nature. It appears strongest 
for non-displaced traps and gets weaker the more the centers of the two traps
are apart from each other. We were also able to calculate the energies of 
the first excited states and interpret the collective motions. For the case of
displaced traps the degeneracies existing in the non-displaced case get lifted.

Acknowledgment: This work was supported by TMR network ERBFMRX-CT96-0002 and
the Austrian Fond zur F\"orderung der wissenschaftlichen Forschung.

% --------------------------------------------------------


\begin{references}


\bibitem{Science95}
M. H. Anderson, J. R. Ensher, M. R. Matthews, C. E. Wieman, E. A. Cornell
Science, {\bf 269}, 198 (1995).

\bibitem{Bradley} 
C. C. Bradley, C. A. Sackett, J. J. 
Tollett, R. G. Hulet, Phys. Rev. Lett. {\bf 75}, 1687 (1995).

\bibitem{Davis} 
K. B. Davis, M.-O. Mewes, M. R. Andrews, N. J. van Druten, D. S. Durfee, 
D. M. Kurn, W. Ketterle, Phys. Rev. Lett. {\bf 75}, 3969 (1995). 

\bibitem{Myatt}
C. J. Myatt, E. A. Burt, R. W. Ghrist, E. A. Cornell, C. E. Wieman,
Phys. Rev. Lett. {\bf 78}, 586 (1997)

\bibitem{Bose} 
S. N. Bose, Z. Phys. {\bf 26}, 178 (1924); 
A. Einstein, Sitz. Preuss Akad. Wiss., 3 (1924).

\bibitem{Edwards}
M. Edwards {\it et al.}, Phys. Rev. A, {\bf 53}, R1950 (1996); Edwards {\it et al.}, preprint;

\bibitem{Stringari}
F. Dalfovo, S. Stringari, Phys. Rev. A {\bf 53} 2477 (1996)
%Bosons in anisotropic traps: Ground state and vortices

\bibitem{Baym}
G. Baym, C. J. Pehtick, Phys. Rev. Lett. {\bf 76} 6 (1996)

\bibitem{Esry}
B. D. Esry, Phys. Rev. A {\bf 55} 1147 (1997)
%one component paper

\bibitem{You}
L. You, W. Hoston, M. Lewenstein, Phys. Rev. A {\bf 55} R1581 (1997)
%Low-energy excitations of trapped Bose condensates

\bibitem{Holland}
M. Holland, D. Jin, M. L. Chiofalo, J. Cooper, preprint

\bibitem{Shly}
Y. Kagan, G. V. Shlyapnikov, J. T. M. Walraven,
Phys. Rev. Lett. {\bf 76}, 2670 (1996)

\bibitem{Castin}
R. Dum, Y. Castin and J. Dalibard, unpublished

\bibitem{Perez}
V. M.  P\'erez-Garc\'{\i}a, H. Michinel, J. I. Cirac, M. Lewenstein, P.
Zoller,
Phys. Rev. Lett. {\bf 77}, 5320 (1996);
V. M.  P\'erez-Garc\'{\i}a, H. Michinel, J. I. Cirac, M. Lewenstein, P.
Zoller,
Phys. Rev. A. (1997) (submitted)

\bibitem{KetterleBoser}
M.-O. Mewes, M. R. Andrews, D. M. Kurn, D. S. Durfee, C. G. Townsend, 
W . Ketterle, Phys. Rev. Lett. {\bf 78} 582 (1997)

\bibitem{KetterleIF}
M. R. Andrews, C. G. Townsend, H.-J. Miesner, D. S. Durfee, D. M. Kurn, 
W. Ketterle, Science {\bf 275} 637 (1997)

\bibitem{Wallis}
H. Wallis, A. Rohrl, M. Naraschewski, A. Schenzle, Phys. Rev. A {55} (2109)
1997
%interference vs. interaction

\bibitem{Burnett}
M. Edwards, P. A. Ruprecht, K. Burnett, R. J. Dodd, C. W. Clark, 
Phys. Rev. Lett. {\bf 77} 1671 (1996)
%Collective excitations of atomic BECs

\bibitem{Ballagh}
R. J. Ballagh, K. Burnett, T. F. Scott, Phys. Rev. Lett. {\bf 78} 1607 (1997)
%Theory of an Output Coupler for Bose-Einstein Condensed Atoms

\bibitem{Stoof}
M. Bijlsma, H. T. C. Stoof, Phys. Rev. A {\bf 55} 498 1997
%Variational approach to the dilute Bose gas

\bibitem{nonlineq}
D. Anderson, 
Phys. Rev. A {\bf 27}, 3135 (1983); 
Z. Fei, V. V. Konotop, M. Peyrard, L. V\'azquez, 
Phys. Rev. E {\bf 48} (1993) 548; 
A. S\'anchez, A. R. Bishop,F. Dom\'{\i}nguez-Adame,
Phys. Rev. E {\bf 49} 4603 (1994) ; 
K. O. Rasmussen, O. Bang, P.L. Christiansen,
Phys. Lett. A {\bf 184} 241 (1994).

\bibitem{Ho}
Tin-Lun Ho, V.B. Shenoy, Phys. Rev. Lett. {\bf 77}, 3276 (1996)

\bibitem{Graham}
R. Graham, D. Walls, {\it preprint} cond-mat/9611111

\bibitem{EsryGreen}
B. D. Esry, C. H. Green, J. P. Burke Jr., J. L. Bohn, preprint;
%two component paper

\bibitem{Desaix}
M. Desaix, D. Anderson, M. Lisak, J. Opt. Soc. Am. B { 8} 2082 (1991)

\bibitem{losses} A phenomenological description of losses can be incorporated
 by assuming complex $U_{nm}=U_{nm}^R + i U_{nm}^I$ ($m=1,2$) with negative
$U_{nm}^I<0$.  The real part  describes elastic collisions
between the particles and the imaginary part accounts for collisional 
losses.

\bibitem{noteexp} In the experiment of Ref. \cite{Myatt} the trap is
anisotropic ($\nu_x=\nu_z=400Hz, \nu_y=11Hz$) and since the spring constants 
of the traps are different for both 
condensate states, the centers of the traps are slightly displaced 
due to gravity.

\bibitem{anothernote} Oscillations in the $y-z$ plane can be expressed as a superposition of $x-y$ and $x-z$  oscillations.



\end{references}
\end{document}